\documentclass[aps,prl,twocolumn]{revtex4-1}

\usepackage{graphicx}% Include figure files
\usepackage{dcolumn}% Align table columns on decimal point
\usepackage{bm}% bold math
\usepackage{braket}
\usepackage{amsmath}
\usepackage{bbold}
\usepackage{verbatim}
\usepackage{hyperref}
\hypersetup{
    colorlinks=true,
    linkcolor=blue,
    filecolor=magenta,      
    urlcolor=cyan,
}

\begin{document}

\title{Diagnosing phase correlations in the joint spectrum of parametric downconversion using multi-photon emission}

\author{Bryn A. Bell$^{1,2}$}
\email{b.bell@imperial.ac.uk} %% email address is required
\author{Gil Triginer Garces$^{1}$}
\author{Ian~A.~Walmsley$^{1,2}$}

\affiliation{1. The Clarendon Laboratory, Department of Physics, University of Oxford, Oxford OX1 3PU, UK.}
\affiliation{2. Quantum Optics and Laser Science Group (QOLS), Department of Physics, Imperial College London, London SW7 2AZ,
UK.}

\preprint{}


\begin{abstract}
The development of new quantum light sources requires robust and convenient methods of characterizing their joint spectral properties. Measuring the joint spectral intensity between a photon pair ignores any correlations in spectral phase which may be responsible for degrading the quality of quantum interference. A fully phase-sensitive characterization tends to require significantly more experimental complexity. Here, we investigate the sensitivity of the frequency-resolved double-pair emission to spectral phase correlations, in particular to the presence of a simple form of correlated phase which can be generated by a chirped pump laser pulse. We observe interference fringes in the four photon coincidences which depend on the frequencies of all four photons, with a period which depends on the strength of their correlation. We also show that phase correlations in the JSA induce spectral intensity correlations between two signal photons, even when the corresponding idler photons are not detected, and link this correlation pattern to the purity of a single signal photon. These effects will be useful in assessing new photon-pair sources for quantum technologies, especially since we require little additional complexity compared to a joint spectral intensity measurement - essentially just the ability to detect at least two photons in each output port.
\end{abstract}
\maketitle
%%%%%%%%%%%%%%%%%%%%%%%%%%  body  %%%%%%%%%%%%%%%%%%%%%%%%%%
\onecolumngrid
The spontaneous parametric downconversion (SPDC) process is ubiquitous in quantum optics, as a source of correlated photon pairs, heralded single photons, entangled photons, or field-quadrature squeezing. Such non-classical states of light have applications in emerging quantum technologies including quantum communication~\cite{gisin07}, quantum metrology~\cite{Giovannetti11}, and optical quantum information processing~\cite{obrien07}. Bringing these technologies to a useful scale places stringent requirements on the performance of sources, which have driven improvements in the brightness and efficiency of SPDC sources~\cite{Eckstein11, Weston16}. Controlling the spectral and temporal properties of the generated light, which can be encapsulated by a joint spectral amplitude (JSA) between a signal and an idler beam, is critical for optimising the quantum features of the twin beams. The JSA can be engineered or affected by the shape of the pump laser pulse and by the properties of the nonlinear medium used: its dispersion and birefringence~\cite{Mosley08}, by placing the medium inside a cavity or resonator~\cite{raymer05,luo15}, or even by using a photonic bandgap structure as the nonlinear medium to suppress emission at certain wavelengths~\cite{Helt17}. Many applications require a factorable JSA (i.e. an absence of correlation or entanglement), such that each photon is emitted into an identical and well-defined spectral-temporal mode, and can undergo high-quality interference with every other photon~\cite{uren05}. The same concerns exist for spontaneous four-wave mixing (FWM) sources, which achieve similar effects as SPDC but using a $\chi^{(3)}$ rather than a $\chi^{(2)}$ nonlinearity, and hence can be implemented in a wider variety of materials~\cite{Rarity05, Dutt15, Spring17}. Progress has also been made in utilising the spectral-temporal mode structure of single photons to encode quantum information~\cite{Brecht15}, with recent demonstrations of heralded single photons in arbitrary temporal shapes and photon-pairs in engineered entangled states~\cite{Ansari18b, Graffiti20}. As a result, increasing attention has been given to methods of characterising the JSA for SPDC and FWM sources, to ensure that a source has the desired spectral-temporal properties, and to diagnose any short-fall in quantum interference visibility.

The most common methods of characterisation are stimulated emission tomography (SET)~\cite{Liscidini13}, where a bright seed laser is used to stimulate a classical analogue of SPDC or FWM, so that the emission can be measured by conventional spectral intensity measurements; or direct measurement at the single photon level, often using a highly dispersive medium as a time-of-flight spectrometer followed by a single photon detector~\cite{Avenhaus09}. In their basic forms, both methods provide the joint spectral intensity (JSI), excluding the phase-information of the JSA, and hence do not constitute a full spectral-temporal characterisation. SET can be made phase-sensitive with an additional phase-reference beam injected into the emission mode~\cite{Jizan16}; similarly, a quantum pulse gate or spectral-shear interferometry can make a direct measurement sensitive to spectral-temporal mode~\cite{Ansari18, Davis18}. All of these techniques require considerable additional experimental complexity in order to retrieve the phase information of the JSA, whether in the form of extra phase-stabilized reference beams, a quantum pulse gate, or an interferometer implementing a spectral shear in one path. Hence for developing quantum photonic technologies with increasing numbers of sources, it is desirable to develop simple and scalable methods of characterisation which are able to diagnose phase-correlations in the JSA which would be missed by a JSI measurement.

In this work, we make spectrally-resolved measurements of photons generated in double-pair emission from a SPDC source. While the frequency-resolved detection of individual photon-pairs only provides information about the JSI, multi-pair emission is sensitive to phase-correlations. This can be understood as interference between different pathways to creating the multiphoton state~\cite{Bell18}: the interfering paths correspond to different ways the photons could be paired with each other, in a manner closely related to Gaussian boson sampling~\cite{Hamilton17}. We operate a type II SPDC source which produces signal and idler beams in horizontal and vertical polarizations. We introduce varying levels of phase-correlation to the JSA, by varying the chirp applied to the pump laser, and observe the resulting interference pattern appearing in the four photon events. We also show that the phase-correlation manifests as spectral intensity correlation between pairs of signal photons, even when the two corresponding idler photons are not detected - this is essentially the spectrally-resolved version of an unheralded $g^{(2)}(0)$ measurement, which is known to be directly related to the purity of the photons~\cite{Christ11}. The spectrally-resolved version provides additional information about the density function of a single photon from a pair, and can provide a more robust estimate of its purity. We note that analogously it has been observed in SET of high-gain sources that the light generated by cascaded or higher-order nonlinear processes is sensitive to phase factors in the first-order SPDC process, though a method of retrieving the full JSA is still lacking~\cite{Triginer19}. 

An ideal two-mode squeezed vacuum (TMSV) state can be written as
\begin{equation}
    \ket{\mathrm{TMSV}}=\sqrt{1-|\xi|^2}~\mathrm{exp}\left(\xi~\hat{a}_s^\dagger \hat{a}_i^\dagger\right)\ket{0,0}=\sqrt{1-|\xi|^2}\sum_{j=0}^\infty \xi^j\ket{j,j},
\end{equation}
with $\ket{j,k}$ a number state of $j$ signal photons and $k$ idler photons, $\hat{a}^\dagger_{s,i}$ the creation operator for signal and idler photons, and $0\leq|\xi|<1$ parameterizing the strength of the squeezing. Introducing spectral amplitude functions for signal and idler $\psi_{s,i}(\omega)$ this becomes
\begin{equation}
    \ket{\mathrm{TMSV}}=\sqrt{1-|\xi|^2}~\mathrm{exp}\left(\xi\iint d\omega_sd\omega_i~\psi_s(\omega_s)\psi_i(\omega_i)~ \hat{a}_s^\dagger(\omega_s)\hat{a}_i^\dagger(\omega_i)\right)\ket{\mathrm{vac}},
\end{equation}
where $\hat{a}_{s,i}^\dagger(\omega)$ is the creation operator for a photon at a specific frequency $\omega$ and $\ket{\mathrm{vac}}$ is the vacuum state over all modes. This is still an ideal TMSV state, with well-defined and separable spectral-temporal modes. A realistic SPDC source will tend to exhibit correlation between signal and idler spectra. Such a multimode squeezed vacuum (MMSV) state can always be written as a product of ideal TMSV states in an orthonormal basis of spectral functions $\psi_{s,i}^{(j)}(\omega)$, known as the Schmidt basis~\cite{Quesada19}, such that
\begin{equation}
    \ket{\mathrm{MMSV}}=\sqrt{\gamma}~\prod_j ~\mathrm{exp}\left(\xi_j\iint d\omega_sd\omega_i~\psi_s^{(j)}(\omega_s)\psi_i^{(j)}(\omega_i)~ \hat{a}_s^\dagger(\omega_s)\hat{a}_i^\dagger(\omega_i)\right)\ket{\mathrm{vac}},
\end{equation}
where $\gamma=\prod_j\left(1-|\xi_j|^2\right)$. Rewriting the product as a sum in the exponent, we arrive at
\begin{equation}
    \ket{\mathrm{MMSV}}=\sqrt{\gamma}~\mathrm{exp}\left(\iint d\omega_sd\omega_i~\psi(\omega_s,\omega_i)~ \hat{a}_s^\dagger(\omega_s)\hat{a}_i^\dagger(\omega_i)\right)\ket{\mathrm{vac}},
\end{equation}
where the JSA function $\psi(\omega_s,\omega_i)=\sum_j\xi_j~\psi_s^{(j)}(\omega_s)\psi_i^{(j)}(\omega_i)$. The probability of generating a pair of photons at particular frequencies is $P(\omega_s,\omega_i)=\gamma|\psi(\omega_s,\omega_i)|^2$. The probability of generating a double pair with additional frequencies $\omega_{s,i}'$ is given by
\begin{equation}
    P(\omega_s,\omega_i,\omega_s',\omega_i')=\gamma\left|\psi(\omega_s,\omega_i)\psi(\omega_s',\omega_i')+\psi(\omega_s,\omega_i')\psi(\omega_s',\omega_i)\right|^2,
    \label{4prob}
\end{equation}
from which it can be seen that there are two paths to creating the four photon event, which add coherently and interfere, such that the four photon probability depends on the relative phases between different points on the JSA.
%Expanding this expression results in two terms which are products of photon-pair probabilities, and hence have no phase dependence, plus an interference term which is sensitive to relative phases between different points on the JSA function:
%\begin{subequations}
%\begin{align}
%    P(\omega_s,\omega_i,\omega_s',\omega_i')&=P_1+P_2+2k\mathrm{Re}(I)\\
%    P_1(\omega_s,\omega_i,\omega_s',\omega_i')&=k^{-1}P(\omega_s,\omega_i)P(\omega_s',\omega_i')\\
%    P_2(\omega_s,\omega_i,\omega_s',\omega_i')&=k^{-1}P(\omega_s,\omega_i')P(\omega_s',\omega_i)\\
%    I(\omega_s,\omega_i,\omega_s',\omega_i')&=\psi(\omega_s,\omega_i)\psi(\omega_s',\omega_i')\psi^*(\omega_s,\omega_i')\psi^*(\omega_s',\omega_i).
%\end{align}
%\end{subequations}

We now consider the effect of applying a chirp to the pump pulse, on a JSA which is otherwise free of correlation. Assuming low parametric gain and low group-velocity dispersion, the JSA has the form
\begin{equation}
    \psi(\omega_s,\omega_i)\propto\alpha(\omega_s+\omega_i)~\phi\left(\frac{n_p-n_s}{c}\omega_s+\frac{n_p-n_i}{c}\omega_i\right)
\end{equation}
with $\alpha(\omega_p)$ the spectral amplitude of the pump pulse and $\phi(\Delta k)$ the phase-matching function, where $n_{p,s,i}$ are the group indices of pump, signal, and idler. Here, $n_s>n_p>n_i$, which allows the JSA to be near-factorable for an appropriate choice of $\alpha(\omega_p)$. We approximate the phase-matching function as Gaussian $\phi(\Delta k)=\mathrm{exp}(-\frac{1}{2}\Delta k^2/\sigma^2)$, in which case the JSA can be made factorable by choosing $\alpha(\omega_p)=\mathrm{exp}(-\frac{1}{2}\omega_p^2/\sigma_p^2)$ with $\sigma_p=c\sigma [(n_s-n_p)(n_p-n_i)]^{-\frac{1}{2}}$. Adding a linear chirp to the pump such that $\alpha(\omega_p)=\mathrm{exp}(-\frac{1}{2}\omega_p^2/\sigma_p^2)~\mathrm{exp}(i\frac{\beta}{2}\omega_p^2)$, we have
\begin{equation}
    \psi(\omega_s,\omega_i)\propto e^{-\frac{1}{2}\omega_s^2/\sigma_s^2}~e^{-\frac{1}{2}\omega_i^2/\sigma_i^2}~e^{i\frac{\beta}{2}(\omega_s+\omega_i)^2}
\end{equation}
where $\sigma_{s,i}=\left(\sigma_p^{-2}+(n_p-n_{s,i})(c\sigma)^{-2}\right)^{-\frac{1}{2}}$. It can be seen that setting $\beta=0$ results in a factorable JSA. Since phase-shifts that are quadratic in either $\omega_s$ or $\omega_i$ do not affect this factorability, we can group them into separate signal and idler functions:
\begin{equation}
    \psi(\omega_s,\omega_i)=\psi_s(\omega_s)\psi_i(\omega_i)e^{i\beta\omega_s\omega_i},
    \label{chirpjsa}
\end{equation}
leaving a non-factorable phase-shift which is bilinear in $\omega_s$ and $\omega_i$, the lowest order of phase correlation. Inserting this form of the JSA into Eq.~\ref{4prob}, the four-photon probability can be written:
\begin{equation}
    P(\omega_s,\omega_i,\omega_s',\omega_i')=4\gamma \Psi_s(\omega_s)\Psi_s(\omega_s')\Psi_i(\omega_i)\Psi_i(\omega_i')~\mathrm{cos}^2\left[\frac{\beta}{2}(\omega_s-\omega_s')(\omega_i-\omega_i')\right],
    \label{fringe4}
\end{equation}
where $\Psi_x(\omega_x)=|\psi_x(\omega_x)|^2$. This probability is separable except for a sinusoidal interference term which depends on all four frequencies and has a modulation period that decreases with increasing $\beta$.

\begin{figure}[t]
	\centering
	\includegraphics[width=0.8\columnwidth]{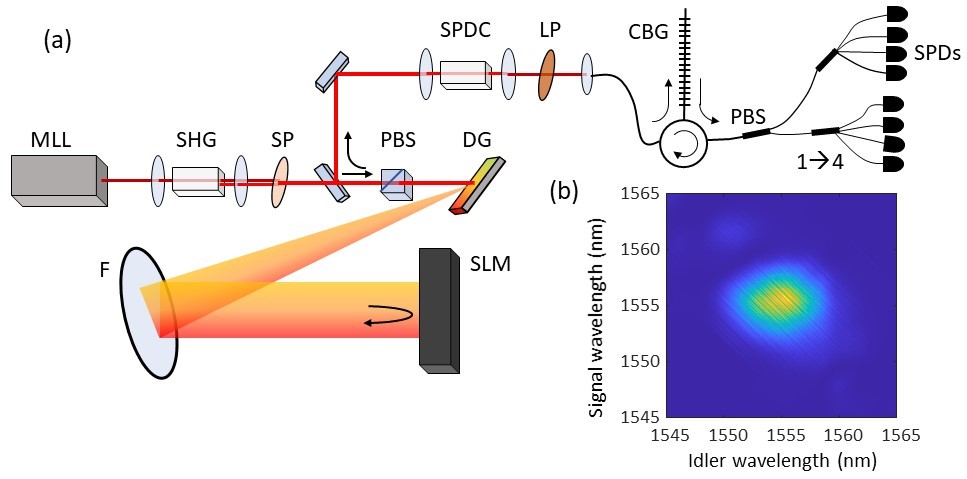}
        \vspace{-10pt}
	\caption{(a) Experimental setup. MLL: mode-locked laser at 1550nm; SHG: second harmonic generation in periodically-poled lithium niobate; SP: short-pass filter; PBS: polarizing beamsplitter; DG: diffraction grating; F: 500mm focusing mirror; SLM: spatial light modulator; SPDC: downconversion in a periodically-poled potassium titanyl phosphate waveguide; LP: long-pass filter; CBG: chirped bragg grating; 1$\rightarrow$4: fiber splitter; SPDs: single photon detectors. (b) Measured JSI for a 2nm pump bandwidth centered at 777nm.}\label{fig1}
 \vspace{-5pt}
\end{figure}

Figure~\ref{fig1}(a) shows the experimental setup to measure this effect. A 10MHz mode-locked fiber-laser produces $\sim100$fs pulses with a broad bandwidth around 1550nm, which are frequency-doubled to $\sim 777$nm with a periodically-poled lithium niobate crystal. A spectral pulse shaper (SPS) is used to set the bandwidth of the pulses and apply a variable chirp, and consists of a spatial light modulator inside a folded 4f line, with a 500mm focal length. The back-reflected pulses from the SPS are picked off and launched into a type II SPDC source based on a highly nonlinear waveguide in periodically-poled potassium titanyl phosphate (ppKTP) crystal, which generates telecom-wavelength near-degenerate photon pairs. The pump is filtered out, while the downconverted photons are collected into fiber. To measure the photons' frequencies, a large group-velocity dispersion is applied using reflection from a chirped fiber Bragg grating and a circulator, which translates wavelength into arrival time with 2.3ns/nm relative delay per change in wavelength. The signal and idler photons have orthogonal polarizations and are separated by a fiber polarizing beamsplitter; then each output is split to four superconducting nanowire single photon detectors, to allow multiple photons to be detected within the detector deadtime. The detector outputs, along with a synchronization signal from the laser, are sent to time-tagging electronics, resulting in an overall $\sim150$ps time resolution, implying $\sim65$pm wavelength resolution for the measured photons. Fig.~\ref{fig1}(b) shows the JSI measured using coincidences between one signal and one idler photon, for a 2nm pump bandwidth, which generates a near-factorable JSI. The phase-matching of the SPDC process is expected to result in a sinc-squared profile containing side-lobes around the central peak: a clear side-lobe can be seen to the top-left of the JSI, but in the bottom-right it appears more smeared out, which is thought to be due to some non-uniformity along the length of the waveguide. Slight diagonal ripples can be seen across the central peak of the JSI, which are caused by ripples in the pump spectrum due to imperfect calibration of the pulse shaper.

\begin{figure}[t]
	\centering
	\includegraphics[width=0.8\columnwidth]{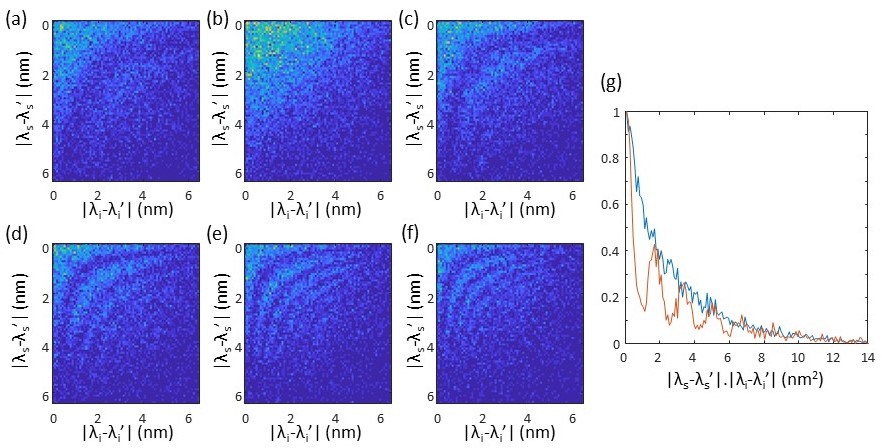}
        \vspace{-10pt}
	\caption{Four photon counts as a function of the difference between the two signal wavelengths and between the two idler wavelengths, for different applied pump chirps: (a) -5ps/nm; (b) 0ps/nm; (c) 5ps/nm; (d) 10ps/nm (e) 15ps/nm; and (f) 20ps/nm. (g) Normalized counts as a function of $|\lambda_s-\lambda_s'|.|\lambda_i-\lambda_i'|$ for applied chirps of 0ps/nm (blue line) and 20ps/nm (orange line).}\label{fig2}
 \vspace{-5pt}
\end{figure}
Figure~\ref{fig2} shows the measured four photon counts, accumulated for 30 minutes, as a function of the difference between the two signal wavelengths, $|\lambda_s-\lambda_s'|$, and the difference between idler wavelengths, $|\lambda_i-\lambda_i'|$. The chirp applied to the pump pulse is varied from -5ps/nm to 20ps/nm from Fig.~\ref{fig2}(a) to (f) in steps of 5ps/nm. The interference is clearly visible as curved fringes, as predicted by Eq.~\ref{fringe4}, which increase in frequency as the chirp, and hence the phase correlation, is increased. These fringes provide a clear signature of the presence of the chirp-induced spectral phase in the two photon joint spectrum. Fig.~\ref{fig2}(g) shows the normalized counts for 0ps/nm and 20ps/nm, as a function of $|\lambda_s-\lambda_s'|.|\lambda_i-\lambda_i'|$. The modulation period of these fringes can be used to infer the value of $\beta$, though they are not dependent on its sign. The contrast of the fringes is limited by residual joint spectral intensity correlations (if the JSI is not separable the two interfering terms in Eq.~\ref{fringe4} will be imbalanced and not cancel fully), background from higher-order events (where more than two pairs are generated), and the finite spectral resolution.

Next we consider the spectral correlations induced between two signal photons by phase correlation in the JSA. If we ignore, i.e. integrate out, the two idler frequencies, from Eq.~\ref{fringe4}, we obtain an expression for the joint probability of two signal photons with frequencies $\omega_s$, $\omega_s'$:
\begin{equation}
    P_s(\omega_s,\omega_s')=2\gamma\Psi_s(\omega_s)\Psi_s(\omega_s')\left[I^2+\left|\tilde{\Psi}_i\left(\beta(\omega_s-\omega_s')\right)\right|^2 \right],
\end{equation}
where $I=\int \Psi_i(\omega_i)d\omega_i$ and $\tilde{\Psi}_x(\tau)$ is the fourier transform of $\Psi_x(\omega_x)$. The constant $I^2$ has no dependence on the phase correlation, whereas the second $\left|\tilde{\Psi}_i\left(\beta(\omega_s-\omega_s')\right)\right|^2$ term does. $\tilde{\Psi}_i\left(\beta(\omega_s-\omega_s')\right)$ is a function centered on the line $\omega_s=\omega_s'$, which becomes progressively narrower as $\beta$ is increased.

For a more general JSA function $\psi(\omega_s,\omega_i)$, i.e. no longer assuming the form given in Eq.~\ref{chirpjsa}, the probability of detecting two signal photons can still be written as the sum of a non-interfering and an interfering term:
\begin{equation}
    P_s(\omega_s,\omega_s')=2\gamma\Psi_s(\omega_s)\Psi_s(\omega_s')+2\gamma|\rho_s(\omega_s,\omega_s')|^2,
\end{equation}
where we redefine $\Psi_s(\omega_s)=\int|\psi(\omega_s,\omega_i)|^2d\omega_i$, and identify the reduced density function for a signal photon, with the corresponding idler traced out, as
\begin{equation}
    \rho_s(\omega_s,\omega_s')=\int \psi(\omega_s,\omega_i)\psi^*(\omega_s',\omega_i)d\omega_i.
\end{equation}
Any partially coherent spectral-temporal shape can be described by such a density function (which is related to a coherence function for bright laser pulses~\cite{Friberg07}), and $\rho_s(\omega_s,\omega_i)$ fully characterizes the signal photons. Here, by subtracting the background non-interfering term, one can experimentally obtain $|\rho_s(\omega_s,\omega_s')|$. Although this is an absolute value and does not contain the full phase information, this does include the absolute values of off-diagonal elements of the density function. These quantify the degree of coherence between two frequencies, and would not be available from an incoherent measurement of the signal photon's spectrum, or from a JSI measurement. %One can see the narrowing of the $\tilde{\Psi}_i\left(\beta(\omega_s-\omega_s')\right)$ function towards the diagonal as the loss of coherence between disparate frequencies.

$|\rho(\omega_s,\omega_s')|^2$ can be measured quickly because it is only necessary to detect two rather than four of the photons; it appears to be relatively sensitive to small amounts of phase correlation; and it is not influenced by loss, even for high levels of squeezing, because with the idler traced out the signal is essentially a multi-mode thermal state.
\begin{figure}[t]
	\centering
	\includegraphics[width=0.8\columnwidth]{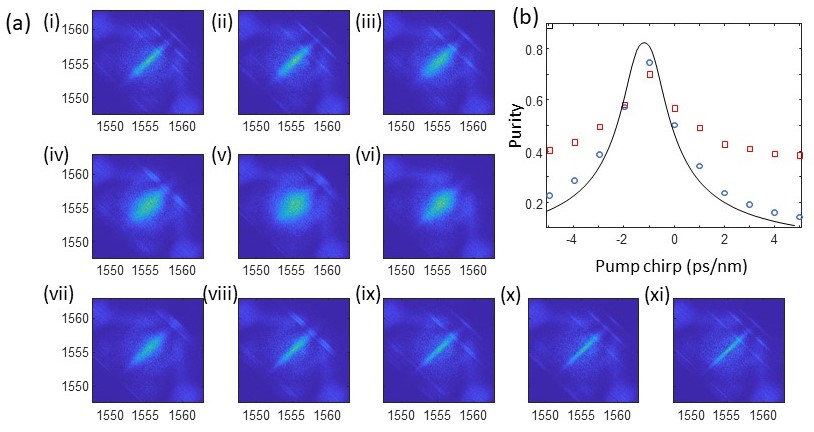}
        \vspace{-10pt}
	\caption{(a) The measured interference term for detection of two signal photons, $I_s(\omega_s,\omega_s')$. (i) to (xi) The chirp applied to the pump pulse is varied from -5ps/nm to +5ps/nm. (b) The estimated purity as a function of pump chirp. black line: simulated value based on the applied chirp; blue circles: value inferred from the narrowing of $I_s(\omega_s,\omega_s')$; red squares: value inferred from the unheralded $g^{(2)}(0)$.}\label{fig3}
 \vspace{-5pt}
\end{figure}
Fig.~\ref{fig3}(a) shows this function measured by detecting two signal photons in coincidence, with the non-interfering term removed by subtracting the number of coincidences measured between separate pump pulses. The chirp applied to the pump was varied from -5ps/nm to +5ps/nmm, in steps of 1ps/nm, and for each value the photon statistics were recorded for 10 minutes. For larger values of chirp, a thin line can be seen along the diagonal, which broadens as expected when the chirp is reduced. This width along the anti-diagonal reaches a maximum at an applied chirp of around -1ps/nm [Fig.~\ref{fig3}(a)(v)], rather than zero as one would expect - this is most likely due to the presence of residual chirp generated by the 4f line in the SPS. The measured phase correlation is minimised when the applied chirp compensates that of the delay line.

The ratio between the total counts in the interfering term $|\rho_s(\omega_s,\omega_s')|^2$ and in the non-interfering term $\Psi_s(\omega_s)\Psi(\omega_s')$ should be equal to the purity of the photons. Since this involves integrating over the remaining frequencies, this does not require any spectral resolution, and is equivalent to an unheralded $g^{(2)}(0)$ measurement, which is commonly used as a test of purity~\cite{Christ11}. However, this value can easily be affected by background, for instance by uncorrelated noise processes. This is particularly problematic when the source is close to degeneracy, because idler photons are more likely to leak into the signal channel, in which case the strong signal-idler correlations artificially inflate the $g^{(2)}(0)$. Fig.~\ref{fig3}(b) shows the expected value of purity from a simulation, based on the shape of the JSI and the applied chirp (black line). The inferred purity from conventional measurements of the unheralded $g^{(2)}(0)$ (red squares) follows broadly the predicted trend, but at larger chirps the value does not fall as far as expected, suggesting this measure is significantly affected by background effects such as leaked idler photons. The blue circles show the purity inferred from the frequency-resolved measurements in Fig.~\ref{fig3}(a), by comparing the narrowed width of the peak along the anti-diagonal to the full width along the diagonal. These values are in much better agreement with the simulation, suggesting that observations from the frequency-resolved measurement can provide a more robust test of purity, avoiding the effects of noise and leaked idler photons.

In conclusion, we have shown that the frequency-resolved double-pair emission from a SPDC source is sensitive to the complex phase of the joint spectral amplitude, and demonstrated that this can be used to detect the presence of a bilinear form of correlated phase. Its effects can be seen as interference fringes in the four photon coincidence probabilities. Further, the coincidences between two signal photons can be used to infer the absolute value of the corresponding single photon density function. This quantifies the degree of coherence between separate frequencies, and can be used to make a robust measure of purity, which is not specific to the bilinear correlated phase. In future work, this could be extended to characterization of more general JSA functions and unknown spectral phase correlations. In analogy with Ref~\cite{Laing12}, where one and two photon interference are used to characterize a linear-optical interferometer, the two and four photon coincidence probabilities should be sufficient for a complete reconstruction of a general JSA, with the proviso that these probabilities are insensitive to non-correlated spectral phases functions affecting the signal or idler individually, as well as to taking the complex conjugate of the whole JSA. Other than these ambiguities, a barrier to this approach is the large number of possible four photon outcomes - for 250 discrete points across the spectral width, around $10^8$ distinct outcomes - which may make it impractical to accumulate sufficient statistics to determine the probability of a particular outcome. A possible way forward is to assume the spectral phase can be modelled using a small number of parameters, e.g. $\beta$ as above and a few describing higher-order phase correlations, and then to retrieve these parameters using maximum likelihood optimization to the photon statistics. Such a method of characterizing the JSA of a photon pair source, including its phase correlations, would be relatively simple to implement, requiring only a time-of-flight spectrometer and the ability to detect at least two photons in each output, and would be valuable in benchmarking new sources and devices for quantum photonic technologies.

\section{Funding}

Engineering and Physical Sciences Research Council (Networked Quantum Information Technologies Hub); European Commission Marie Sk\l{}odowska Curie Individual Fellowship (FrEQuMP, 846073).

%\section*{Acknowledgment}

%Acknowledgments

%Manual citation list


\begin{thebibliography}{1}
\bibitem{gisin07}
N.~Gisin and R.~Thew, ``Quantum Communication,'' Nature Photonics \textbf{1}, 165-171 (2007).

\bibitem{Giovannetti11}
V.~Giovannetti, S.~Lloyd, and L.~Maccone, ``Advances in quantum metrology,'' Nature Photonics \textbf{5}, 222–229 (2011).

\bibitem{obrien07}
J.L.~O'Brien, ``Optical quantum computing,'' Science \textbf{318}, 1567-1570 (2007).

\bibitem{Eckstein11}
A.~Eckstein, A.~Christ, P.J.~Mosley, and C.~Silberhorn, ``Highly Efficient Single-Pass Source of Pulsed Single-Mode Twin Beams of Light,'' Phys. Rev. Lett. \textbf{106}, 013603 (2011).

\bibitem{Weston16}
M.M.~Weston, H.M.~Chrzanowski, S.~Wollmann, A.~Boston, J.~Ho, L.K.~Shalm, V.B.~Verma, M.S.~Allman, S.W.~Nam, R.B.~Patel, S.~Slussarenko, and G.J.~Pryde, ``Efficient and pure femtosecond-pulse-length source of polarization-entangled photons,'' Opt. Express \textbf{24}, 10869-10879 (2016).

\bibitem{Mosley08}
P.J.~Mosley, J.S.~Lundeen, B.J.~Smith, P.~Wasylczyk, A.B.~U'Ren, C.~Silberhorn, I.A.~Walmsley, ``Heralded generation of ultrafast single photons in pure quantum states,'' Phys. Rev. Lett. \textbf{100}, 133601 (2008).

\bibitem{raymer05}
M.G.~Raymer, J.~Noh, K.~Banaszek, and I.A.~Walmsley, ``Pure-state single-photon wave-packet generation by parametric down-conversion in a distributed microcavity,'' Phys. Rev. A \textbf{72}, 023825 (2005).

\bibitem{luo15}
K.-H.~Luo, H.~Herrmann, S.~Krapick, B.~Brecht, R.~Ricken, V.~Quiring, H.~Suche, W.~Sohler and C.~Silberhorn, ``Direct generation of genuine single-longitudinal-mode narrowband photon pairs,'' New J. Phys. \textbf{17}, 073039 (2015).

\bibitem{Helt17}
L.G.~Helt, A.M.~Bra\'{n}czyk, M.~Liscidini, M.J.~Steel, ``Parasitic Photon-Pair Suppression via Photonic Stop-Band Engineering,'' Phys. Rev. Lett. \textbf{118}, 073603 (2017).

\bibitem{uren05}
A.B.~U'Ren, C.~Silberhorn, R.~Erdmann, K.~Banaszek, W.P.~Grice, I.A.~Walmsley, M.G.~Raymer, ``Generation of Pure-State Single-Photon Wavepackets by Conditional Preparation Based on Spontaneous Parametric Downconversion,'' Las. Phys. \textbf{15} 146 (2005).

\bibitem{Rarity05}
J.G.~Rarity, J.~Fulconis, J.~Duligall, W.J.~Wadsworth,
and P.S.J.~Russell, ``Photonic crystal fiber source of correlated photon pairs,'' Opt. Express \textbf{13}, 534 (2005).

\bibitem{Dutt15}
A.~Dutt, K.~Luke, S.~Manipatruni, A.L.~Gaeta, P.~Nussenzveig, and M.~Lipson, ``On-Chip Optical Squeezing,'' Phys. Rev. Applied \textbf{3}, 044005 (2015)

\bibitem{Spring17}
J.B.~Spring, P.L.~Mennea, B.J.~Metcalf, P.C.~Humphreys, J.C.~Gates, H.L.~Rogers, C.~S\"{o}ller, B.J.~Smith, W.S.~Kolthammer, P.G.R.~Smith, and I.A.~Walmsley, ``Chip-based array of near-identical, pure, heralded single-photon sources,'' Optica \textbf{4}, 90-96 (2017).

\bibitem{Brecht15}
B.~Brecht, D.V.~Reddy, C.~Silberhorn, and M.G.~Raymer, ``Photon Temporal Modes: A Complete Framework for Quantum Information Science,'' Phys. Rev. X \textbf{5}, 041017 (2015).

\bibitem{Ansari18b}
V.~Ansari, E.~Roccia, M.~Santandrea, M.~Doostdar, C.~Eigner, L.~Padberg, I.~Gianani, M.~Sbroscia, J.M.~Donohue, L.~Mancino, M.~Barbieri, and C.~Silberhorn, ``Heralded generation of high-purity ultrashort single photons in programmable temporal shapes,'' Opt. Express \textbf{26}, 2764-2774 (2018).

\bibitem{Graffiti20}
F.~Graffitti, P.~Barrow, A.~Pickston, A.M.~Bra\'{n}czyk, A.~Fedrizzi, ``Direct generation of tailored pulse-mode entanglement,'' Phys. Rev. Lett. \textbf{124}, 053603 (2020).

\bibitem{Liscidini13}
 M.~Liscidini and J.~Sipe, ``Stimulated emission tomography'', Phys. Rev. Lett. \textbf{111}, 193602 (2013).
 
 \bibitem{Avenhaus09}
 M.~Avenhaus, A.~Eckstein, P.J.~Mosley, and C.~Silberhorn, ``Fiber-assisted single-photon spectrograph,'' Opt. Lett.
\textbf{34}, 2873–2875 (2009).

\bibitem{Jizan16}
I.~Jizan, B.~Bell, L.G.~Helt, A.~Casas Bedoya, C.~Xiong, B.J.~Eggleton, ``Phase-sensitive tomography of the joint spectral amplitude of photon pair sources,'' Opt. Lett. \textbf{41}, 4803-4806 (2016).

\bibitem{Ansari18}
V.~Ansari, J.M.~Donohue, M.~Allgaier, L.~Sansoni, B.~Brecht, J.~Roslund, N.~Treps, G.~Harder, C.~Silberhorn, ``Tomography and purification of the temporal-mode structure of quantum light,'' Phys. Rev. Lett. \textbf{120}, 213601 (2018).

\bibitem{Davis18}
A.O.C.~Davis, V.~Thiel, M.~Karpi\'{n}ski, and B.J.~Smith, ``Measuring the single-photon temporal-spectral wave function,'' Phys. Rev. Lett. \textbf{121}, 083602 (2018).

\bibitem{Triginer19}
G.~Triginer, M.D.~Vidrighin, N.~Quesada, A.~Eckstein,
M.~Moore, W.S.~Kolthammer, J.E.~Sipe, I.A.~Walmsley, ``Understanding high gain twin beam sources
using cascaded stimulated emission,'' arXiv:1911.05708v2 [quant-ph] (2019).

\bibitem{Quesada19}
N.~Quesada, G.~Triginer, M.D.~Vidrighin, J.E.~Sipe, ``Efficient simulation of high-gain twin-beam generation in waveguides,'' arXiv:1907.01958v2 [quant-ph] (2019).

\bibitem{Bell18}
B.A.~Bell and B.J.Eggleton, ``Multiphoton interference in the spectral domain by direct heralding of frequency superposition states,'' Phys. Rev. Lett. \textbf{121}, 033601 (2018).

\bibitem{Hamilton17}
C.S.~Hamilton, R.~Kruse, L.~Sansoni, S.~Barkhofen, C.~Silberhorn, and I.~Jex, ``Gaussian boson sampling,'' Phys. Rev. Lett. \textbf{119}, 170501 (2017).

\bibitem{Christ11}
A.~Christ, K.~Laiho, A.~Eckstein, K.N.~Cassemiro, C.~Silberhorn, ``Probing multimode squeezing with correlation functions,'' New J. Phys. \textbf{13}, 033027 (2011).

\bibitem{Friberg07}
A.T.~Friberg, H.~Lajunen, V.~Torres-Company, ``Spectral elementary-coherence-function representation for partially coherent light pulses,'' Opt. Express \textbf{15}, 5160-5165 (2007).

\bibitem{Laing12}
A.~Laing and J.L.~O'Brien, ``Super-stable tomography of any linear optical device,'' arXiv:1208.2868 [quant-ph] (2012).

\end{thebibliography}
\end{document}